\documentstyle[prl,aps,multicol,psfig,array]{revtex}

\newcommand{\beq}{\begin{equation}}
\newcommand{\eeq}{\end{equation}}
\newcommand{\bqa}{\begin{eqnarray}}
\newcommand{\eqa}{\end{eqnarray}}

\def\square{\vcenter{\vbox{\hrule height.4pt
          \hbox{\vrule width.4pt height8pt
          \kern8pt\vrule width.4pt}\hrule height.4pt}}}

\begin{document}

\preprint{
\vbox{\halign{&##\hfil\cr
&ITF-UU-01/29\cr        & cond-mat/yymmnn \cr
}}}

\title{Ground State Pressure and Energy Density of a 
 Homogeneous Bose Gas in Two Dimensions}

\author{Jens O. Andersen}
\address{Institute for Theoretical Physics, University of Utrecht,\\
       Leuvenlaan 4, 3584 CE Utrecht, The Netherlands\\
(\today)}

\maketitle

\begin{abstract}
{
We consider an interacting homogeneous 
Bose gas at zero temperature in two spatial dimensions.  
The properties of the system can be calculated as an expansion in powers of
$g$, where $g$ is the coupling constant. 
We calculate the ground state
pressure and the ground state
energy density to second order in the quantum loop expansion.
The renormalization group
is used to sum up leading and subleading logarithms from all orders
in perturbation theory.
In the dilute limit, the renormalization group improved 
pressure and energy density are expansions in powers
of the $T$-matrix.
}
\end{abstract}

\begin{multicols}{2}
\section{Introduction}
The remarkable achievement of Bose-Einstein condensation (BEC) of alkali
atoms in harmonic traps~\cite{bec1,bec2,bec3} 
has created an enormous interest in the properties of dilute Bose gases
(For a review, see e.g. Ref.~\cite{string} and references therein).

The homogeneous Bose gas 
in three dimensions has been studied in
great detail over the past 50 years (For a review, see e.g.~\cite{grif}). 
At zero temperature,
the quantum loop expansion is essentially an expansion in powers
of $\sqrt{\rho a^3}$, where $a$ is the two-body $S$-wave scattering length
and $\rho$ is the density.
Lee and Yang~\cite{leeyang} 
were the first to calculate the leading
quantum correction to the energy density. 
Part of the second quantum correction to the energy density
was obtained by Wu, by 
Hugenholz and Pines, and by Sawada~\cite{to}.
Only recently a complete two-loop result has been 
obtained by Braaten and Nieto~\cite{eric}.
The result depends, in addition to the scattering length, also
on the scattering amplitude for $3\rightarrow 3$ scattering.

The homogeneous Bose gas in two dimensions has also been studied extensively.
The chemical potential and 
ground state energy density of a two-dimensional homogeneous Bose gas
were first calculated by Schick~\cite{schiff}.
By summing up the ladder diagrams contributing to the
chemical potential, he was able to show that, in the dilute limit, the 
relevant expansion parameter is not the coupling constant $g$, but
rather the effective interaction
$[\log(\rho a^2)]^{-1}$, where $\rho$ is the density and $a$
is the range of the interaction. The expansion parameter is the 
two-body $T$-matrix, and 
Schick determined the leading-order results in this expansion.
Corrections to these results have been considered by several 
authors~\cite{popov1,popov2,fis,rg2,fried,hines}.
Very recently, a formal proof of the result by Schick was given by
Lieb and Yngvason~\cite{lieb1,lieb2}. 

In the present paper, we reconsider the homogeneous Bose gas in two
dimensions at zero temperature. We calculate the
pressure and 
energy density of the ground state to second order in the quantum
loop expansion. We also apply the renormalization group to sum up leading
and subleading logarithms from all orders of perturbation theory.
In the dilute limit, the renormalization group improved 
pressure and energy density are essentially expansions in powers
of the $T$-matrix.

The outline of the paper is as follows. In Sec.~II, we 
briefly discuss the perturbative framework developed in Ref.~\cite{eric}
to calculate the ground state properties of a homogeneous Bose gas.
In Sec.~III, we calculate the ground state pressure
to two loop order.
In Sec.~IV, we 
calculate the ground state
energy density through two loops. 
In Sec.~V, we apply the renormalization group to sum up leading
and subleading logarithms from all orders in perturbation theory.
Finally, we summarize our results in Sec.~VI.
Calculational details are included in an appendix.

\section{Perturbative Framework}
In this section, we discuss the perturbative framework set up in 
Ref.~\cite{eric} to calculate the
effects on the ground state
from quantum fluctuations around the mean field.

The action is
\bqa\nonumber
S&=&\int dt\;
\Bigg\{
\int d^2x\;
\psi^*({\bf x},t)
\!\!\left[
i\hbar{\partial\over\partial t}
+{\hbar^2\over2m}\nabla^2+\mu
\right]
\psi({\bf x},t)
\\ \nonumber
&&
\hspace{-.2cm}
-{1\over2}\int d^2x\int d^2x^{\prime}\;
\Bigg[
\psi^*({\bf x},t)\psi^*({\bf x}^{\prime},t) V_0({\bf x}-{\bf x}^{\prime})
\times
\\
&&
\psi({\bf x},t)\psi({\bf x}^{\prime},t)\Bigg]
\Bigg\}
\;.
\label{ac}
\eqa
$\psi^*({\bf x},t)$ is a complex field operator that creates a boson at the
position ${\bf x}$, $\mu$ is the chemical potential, and $V_0({\bf x})$ 
is the two-body potential. In the following, we set $\hbar=2m=1$.
Factors of $\hbar$ and $2m$ can be reinserted using dimensional analysis.

The action Eq.~(\ref{ac}) is symmetric under a phase transformation
\bqa
\psi({\bf x},t)\rightarrow e^{i\alpha}\psi({\bf x},t)\;.
\eqa

The $U(1)$-symmetry ensures that the density $\rho$ and current 
density ${\bf j}$ satisfy the continuity equation
\bqa
\dot{\rho}+\nabla\cdot{\bf j}=0\;.
\eqa
In the ground state, the current density ${\bf j}$ vanishes identically and
the condensate has a constant phase. The $U(1)$-symmetry can then be used to
make the condensate real everywhere.

If the interatomic potential $V_0({\bf x})$ is short range, 
it can be mimicked by local interactions. If the 
energies are low enough, the scattering amplitude 
can be approximated by $s$-wave scattering and the action Eq.~(\ref{act}) can
be approximated by~\cite{e2} 
\bqa
S&=&\int dt
\int d^2x\;
\psi^*
\left[
i
{\partial\over\partial t}
+\nabla^2+\mu
\right]\psi
-{1\over2}
g\Big(\psi^*\psi\Big)^2
\;.
\label{act}
\eqa
Here, $g$ is a coupling constant that must be tuned to reproduce 
some low-energy observable of the true potential $V_0({\bf x})$.

The quantum field theory defined by the action Eq.~(\ref{act})
has ultraviolet divergences that must be removed by renormalization of
$\mu$ and $g$. 
There is also an ultraviolet divergence in the expression for the density
$\rho$. This divergence can be removed by adding a counterterm $\delta\rho$.
Alternatively, one can eliminate the divergences associated with
$\mu$ and $\rho$ by a normal-ordering prescription of the fields 
in Eq.~(\ref{act}). The coupling constant is renormalized in the usual way
by replacing the bare coupling with the physical one.

If we use a simple momentum cutoff $M$ to cut off the
ultraviolet 
divergences in the loop integrals, there will be terms proportional to
$M^p$, where $p$ is a positive integer.
There are also terms that are proportional to $\log M$.
The coefficients of the power divergences depend on the regularization
method and are therefore artifacts of the regulator.
On the other hand, the coefficients of $\log(M)$ are independent of the
regulator and they therefore represent real physics.
In this paper, we use dimensional regularization 
to regulate both infrared and ultraviolet divergences.
In dimensional regularization, one calculates the loop
integrals in $d=2-2\epsilon$ dimensions for values of $\epsilon$ where
the integrals converge. One then analytically continues back to $d=2$
dimensions.
With dimensional regularization, an arbitrary renormalization scale $M$
is introduced. This scale can be identified with the simple momentum
cutoff mentioned above.
An advantage of dimensional regularization is that it 
automatically sets power 
divergences to zero, while logarithmic divergences
show up as poles in $\epsilon$.
In two dimensions, the one-loop 
counterterms for the chemical potential $\mu$ and the density $\rho$
are quadratic ultraviolet
divergences, while the one-loop counterterm for the coupling constant
$g$ is a logarithmic ultraviolet divergence.
At the two-loop level, the counterterms for the chemical potential and 
the density are also quadratic divergences. The counterterm for the 
coupling constant is a double logarithmic divergence.

We next parameterize the quantum field $\psi$
in terms of a time-independent
condensate $v$ and a quantum fluctuating field $\tilde{\psi}$:
\bqa
\psi=v+\tilde{\psi}\;.
\eqa
The fluctuating field can be written in terms of two real fields:
\bqa
\label{split}
\tilde{\psi}={1\over\sqrt{2}}\left(\psi_1+i\psi_2\right)\;.
\eqa
Substituting Eq.~(\ref{split}) into Eq.~(\ref{act}), the action can 
be decomposed into three terms
\bqa
\label{terms}
S[\psi]=S[v]+S_{\rm free}[\psi_1,\psi_2]
+S_{\rm int}[v,\psi_1,\psi_2]\;.
\eqa
$S[v]$ is the classical action
\bqa
S[v]=\int dt\int d^2x
\left[\mu v^2-{1\over2}gv^4
\right]\;,
\eqa
while the free part of the action is
\bqa\nonumber
S_{\rm free}[\psi_1,\psi_2]&=&\int dt\int d^2x
\left[
{1\over2}\left(\dot{\psi}_1\psi_2-\psi_1\dot{\psi_2}\right)
\right.\\
&&
\hspace{-1cm}
\left.
+{1\over2}
\psi_1\left(\nabla^2+X\right)\psi_1+{1\over2}
\psi_2\left(
\nabla^2+Y\right)\psi_2
\right]\;.
\label{free}
\eqa
The interaction part of the action is
\bqa\nonumber
S_{\rm int}[v,\psi_1,\psi_2]&=&
\int dt\int d^2x\left[
\sqrt{2}T\psi_1
\right. \\
&&
\hspace{-1cm}
\left.
+{1\over\sqrt{2}}Z\psi_1\left(\psi_1^2+\psi_2^2\right)
-{1\over8}g\left(\psi_1^2+\psi_2^2\right)^2
\right]\;.
\label{inter}
\eqa
The sources in Eq.~(\ref{inter}) are
\bqa
T&=&\left[\mu-gv^2\right]v
\;,\\
X&=&
\left[\mu-3gv^2\right]\;,\\
Y&=&
\left[\mu-gv^2\right]\;,\\
Z&=&-gv\;.
\eqa
The propagator that corresponds to the free action $S_{\rm free}
[\psi_1,\psi_2]$
in Eq.~(\ref{free}) is
\bqa
\label{prop}
D(\omega,p)&=&\frac{i}{\omega^2
-\epsilon^2(p)+i\epsilon}\left(\begin{array}{cc}
p^2-Y&-i\omega \\
i\omega&p^2-X
\end{array}\right)\;.
\eqa
Here ${\bf p}$ is the wavevector, $\omega$ is the frequency, and
$\epsilon(p)$ is the dispersion relation:
\bqa
\label{disp}
\epsilon(p)=\sqrt{(p^2-X)(p^2-Y)}\;.
\eqa
The value of the condensate $v_0$ that minimizes the classical action is given
by the equation $T=0$.
Both the propagator Eq.~(\ref{prop}) and the dispersion relation 
Eq.~(\ref{disp}) greatly simplify at $v_0$, since
$Y=0$ here.
The dispersion relation then becomes 
gapless, which 
reflects the spontaneous breakdown of the $U(1)$-symmetry
(Goldstone's theorem). 
The dispersion relation is
linear for small wavevectors and 
is quadratic for large wavevectors, which is that of a 
free nonrelativistic 
particle. The propagator is defined with an $i\epsilon$ prescription
in the usual way. The diagonal parts of the propagator are denoted
by a solid and a dashed line, respectively. The off-diagonal parts
are denoted by a solid-dashed and dashed-solid line, respectively.

The partition function ${\cal Z}$ can be expressed as a path integral
\bqa
{\cal Z}=\int{\cal D}\psi_1{\cal D}\psi_2\,e^{iS[\psi_1,\psi_2]}\;.
\eqa
All the thermodynamic observables can be derived from ${\cal Z}$.
For instance, the free energy density ${\cal F}$ is given by
\bqa
{\cal F}(\mu)=i{\log{\cal Z}\over VT}\;,
\eqa
where $VT$ is the spacetime volume of the system.

The density $\rho$ is given by the expectation value 
$\langle\psi^{\dagger}\psi\rangle$ in the ground state. 
It can therefore be expressed
as 
\bqa
\label{rhodef}
\rho(\mu)=-{\partial {\cal F}(\mu)\over \partial \mu}\;.
\eqa
The energy density ${\cal E}$ is given by the Legendre transform of the
the free energy density
\bqa
\label{rela}
{\cal E}(\rho)={\cal F}(\mu)+\rho\mu\;.
\eqa

At this point it is convenient to introduce the 
thermodynamic potential $\Omega(\mu,v)$.  
The thermodynamic potential is given by all one-particle 
irreducible vacuum graphs and 
can be expanded in the number of loops
\bqa
\label{loopp}
\Omega(\mu,v)&=&\Omega_0(\mu,v)+\Omega_1(\mu,v)+\Omega_2(\mu,v)+...\;,
\eqa
where the subscript $n$ indicates the contribution from the
nth order in the loop expansion.
The free energy ${\cal F}$ is given by 
all connected vacuum graphs and is independent of the 
condensate $v$. If we evaluate $\Omega$ at a value of the condensate that
satisfies the condition
\bqa
\label{vc}
\bar{v}=\langle\psi\rangle\;,
\eqa
it can be shown that all the one-particle reducible graphs (those that can
be disconnected by cutting a single line) vanish.
We then have
\bqa
\label{loop}
{\cal F}(\mu)&=&
\Omega_0(\mu,\bar{v})+\Omega_1(\mu,\bar{v})+\Omega_2(\mu,\bar{v})+...\;.
\eqa
The condition Eq.~(\ref{vc}) is equivalent to 
\bqa
\label{tad1}
{\partial\Omega\over \partial v}=0\;.
\eqa
The loop expansion Eq.~(\ref{loop}) does not coincide with the expansion
in powers quantum corrections. To obtain the expansion in powers of
quantum correction, we must
expand the condensate $\bar{v}$ about the classical minimum $v_0$:
\bqa
\bar{v}=v_0+v_1+v_2+...\;.
\eqa
By substituting Eq.~(\ref{loopp}) into Eq.~(\ref{tad1}), we obtain the
first quantum correction $v_1$ to the classical minimum: 
\bqa
\label{v1}
v_1=-{\partial\Omega_1(\mu,v)\over\partial v}\Bigg|_{v=v_0}\Bigg/
{\partial^2\Omega_0(\mu,v)\over\partial v^2}\Bigg|_{v=v_0}
\;.
\eqa
This first quantum correction to the free energy density is
\bqa
{\cal F}_1(\mu)=\Omega_1(\mu,v_0)\;,
\eqa
and the second quantum correction to the free energy density
is
\bqa\nonumber
{\cal F}_2(\mu)&=&\Omega_2(\mu,v_0)+
v_1{\partial \Omega_1(\mu,v)\over\partial v}\Bigg|_{v=v_0}
\\
&&
+{1\over2}v^2_1{\partial^2\Omega_0(\mu,v)\over\partial v^2}\Bigg|_{v=v_0}\;.
\eqa

\section{Pressure to two loops}
In this section, we calculate the pressure as a function of the
chemical potential $\mu$ and the renormalized coupling $g$ to two loops.
\subsection{Mean-field free energy}
The thermodynamic potential in the mean-field approximation is
\bqa
\Omega_0(\mu,v)=-
\mu v^2+{1\over2}gv^4\;.
\eqa
The mean-field free energy is given by 
the classical thermodynamic potential evaluated at the classical
minimum $v_0$:
\bqa
\label{f0}
{\cal F}_0(\mu)=-{\mu^2\over2g}\;.
\eqa
\subsection{One-loop free energy}
The one-loop contribution to the free energy is 
\bqa
\label{f111}
{\cal F}_{1}(\mu)=
{\cal F}_{1a}(\mu)+\Delta_1{\cal F}\;,
\eqa
where
\bqa
\label{f1a}
{\cal F}_{1a}(\mu)=
{i\over2}\int{d\omega\over 2\pi}\int{d^dp\over(2\pi)^d}\log\det D^{-1}\;,
\eqa
and $\Delta_1{\cal F}$ is the one-loop counterterm
\bqa
\label{1c}
\Delta_1{\cal F}&=&-{\mu\over g}\Delta_1\mu+{\mu^2\over2g^2}\Delta_1g\;.
\eqa
The propagator $D(\omega,p)$ is evaluated at the classical minimum, where
$Y=0$.
By integrating over $\omega$, Eq.~(\ref{f1a}) becomes
\bqa\nonumber
{\cal F}_{1a}(\mu)&=&{1\over2}\int{d^dp\over(2\pi)^d}\;\epsilon(p)\;\\
\label{f1div}
&=&{1\over2}I_{0,-1}(2\mu)\;,
\eqa
where $I_{m,n}$ is defined in the appendix.
Using Eq.~(\ref{i1}) in the appendix, we obtain 
\bqa
\label{divv}
{\cal F}_{1a}(\mu)=-{\mu^2\over16\pi}\left[{1\over\epsilon}
-L-{1\over2}
+{\cal O}(\epsilon)
\right]\;,
\eqa
where $L=\log\left(\mu/2M^2\right)$.
The counterterms at one-loop are~\cite{berg}:
\bqa
\label{dm1}
\Delta_1\mu&=&0\;, \\
\label{dg1}
\Delta_1g&=&{g^2\over8\pi\epsilon}\;.
\eqa
Adding Eqs.~(\ref{1c}) and~(\ref{divv}), we obtain
the one-loop contribution to the free energy
\bqa
{\cal F}_{1}(\mu)=
{\mu^2\over16\pi}\left[L+{1\over2}\right]\;.
\eqa 
Note that ${\cal F}_1(\mu)$ is independent of the coupling
constant $g$.
Adding Eqs.~(\ref{f0}) and~(\ref{f1}), we obtain the one-loop approximation to
the free energy:
\bqa
\label{f1}
{\cal F}_{0+1}(\mu)=
-{\mu^2\over g}+
{\mu^2\over16\pi}\left[L+{1\over2}\right]\;,
\eqa
where the coupling constant is evaluated at the scale $M$, $g=g(M)$.
\subsection{Two-loop free energy}
\vspace{-1.5cm}
\begin{figure}
\psfig{figure=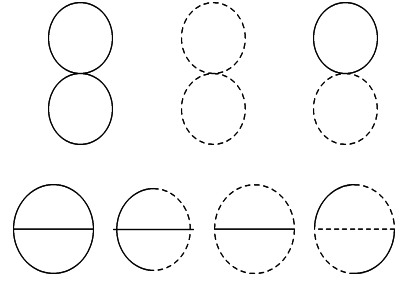,width=9cm,height=14cm}
\vspace{-7.5cm}
\caption{\narrowtext
Two-loop vacuum graphs contributing to the free energy ${\cal F}$.}
\label{2loop}
\end{figure}

The two-loop contribution to the thermodynamic potential $\Omega_2(\mu,v)$
isgiven by the one-particle irreducible graphs
shown in Fig.~\ref{2loop}. The expression for the diagrams is
\bqa\nonumber
\Omega_2(\mu,v)&=&{g\mu\over8}J
+{g\over32}
\left[3I_{1,1}^2+2I_{-1,-1}I_{1,1}
\right.
\\&&
\left.
+3I_{-1,-1}^2\right]\;,
\eqa
where 
\bqa
J=6J_{0,0,1}-J_{-1,-1,1}-3J_{1,1,1}-2J_{-1,0,0}\;, 
\eqa
and
integrals $J_{l,m,n}$ are defined in the appendix.
The first correction $v_1$ is given by Eq.~(\ref{v1}):
\bqa
v_1=-{gv_0\over8\mu}\left[3I_{1,1}+I_{-1,-1}\right]\;.
\eqa
The two-loop contribution to the free energy is then 
\bqa\nonumber
{\cal F}_2(\mu)&=&
{g\mu\over8}J
+{g\over16}
\left[I_{-1,-1}^2-2I_{-1,-1}I_{1,1}-3I_{1,1}^2\right]
\\
&&
\label{f2j}
+\Delta_2{\cal F}\;.
\eqa
The counterterm $\Delta_2{\cal F}$ is
\bqa\nonumber
\Delta_2{\cal F}&=&
{1\over2}I_{1,1}\Delta_1\mu
-{\mu\over g}\Delta_2\mu+{\mu^2\over2g^2}\Delta_2g
\\
&&
-{1\over2g}\left(\Delta_1\mu\right)^2+{\mu\over g^2}\Delta_1\mu\Delta_1g
-{\mu^2\over2g^3}\Delta_1g^2
\;.
\eqa
At the two-loop level, the counterterms are given by
\bqa
\label{dm2}
\Delta_2\mu&=&0\;, \\
\label{dg2}
\Delta_2g&=&{g^3\over64\pi^2\epsilon^2}\;.
\eqa
The integrals $J_{l,m,n}$ are ultraviolet divergent. After subtracting the
divergent part, the remainder must be calculated numerically. We evaluate
the expression in the appendix. The result is
\bqa
J=-2I_{0,1}\left[I_{1,1}+I_{-1,-1}\right]+J_{\rm num}\;,
\eqa
where $J_{\rm num}=-3.52\times10^{-5}\mu$.
The final result for the two-loop contribution to the free energy is
\bqa
\label{f2}
{\cal F}_2(\mu)&=&-{g\mu^2\over64\pi^2}\left[1+C\right]
\;.
\eqa 
Here $C=2.78\times10^{-3}$.
Adding Eqs.~(\ref{f0}),~(\ref{f1}), and~(\ref{f2}), 
we obtain our final result for the free energy to second order
in the quantum loop expansion
\bqa
\label{f012}
{\cal F}_{0+1+2}(\mu)&=&
-{\mu^2\over2g}+{\mu^2\over16\pi}\left[L+{1\over2}\right]
-{g\mu^2\over64\pi^2}\left[1+C\right]
\;,
\eqa
\subsection{Pressure to two loops}
The pressure ${\cal P}$ is given by $-{\cal F}$. The pressure through
two loops is given by minus the sum of 
Eqs.~({\ref{f0}}),~(\ref{f1}) and~(\ref{f2}):
\bqa
\label{pres2}
{\cal P}_{0+1+2}(\mu)={\mu^2\over2g}-{\mu^2\over16\pi}\left[L+{1\over2}\right]
+{g\mu^2\over64\pi^2}\left[1+C\right]
\;.
\eqa
The coupling constant $g$ in Eq.~(\ref{act}) 
satisfies
\bqa
\label{rg2}
M{\!\!d\over dM}\;g&=&\beta(g)\;,
\eqa
where the $\beta$-function is a polynomial in $g$.
Normally, the $\beta$-functions are
known only up to a certain order in the quantum loop
expansion. In the present case, the one-loop result for 
the $\beta$-function is exact
and $\beta(g)=g^2/4\pi$~\cite{berg}. 
From Eq.~(\ref{rg2}), one can easily
check that our result
Eq.~(\ref{pres2}) for the two-loop pressure is independent of the scale $M$
up to correction of order $g^2$.
\section{Energy density to two loops}
In this section, we derive the energy density ${\cal E}$ 
as a function of the density $\rho$ and the renormalized coupling
$g$ to two loops.
\subsection{Mean field energy density}
Using Eqs.~(\ref{rhodef}) and~(\ref{f0}) the density in the mean-field 
approximation is 
\bqa
\label{mu0}
\rho_{0}(\mu)={\mu\over g}\;.
\eqa
The chemical potential is obtained by inverting Eq.~(\ref{mu0}): 
\bqa
\mu_{0}(\rho)&=&g\rho\;.
\eqa
Using Eqs.~(\ref{rela}) and~(\ref{mu0}), the energy density 
in the mean-field approximation is given by
\bqa
\label{e0}
{\cal E}_0(\rho)&=&{1\over2}g\rho^2\;.
\eqa
\subsection{One-loop energy density}
Using Eqs.~(\ref{rhodef}) and~(\ref{f1})  
we obtain the density in the
one-loop approximation
\bqa
\label{rho1}
\rho_{0+1}(\mu)=
{\mu\over g}-{\mu\over8\pi}\bigg[L+1\bigg]\;.
\eqa
Inverting Eq.~(\ref{rho1}) to obtain $\mu$ as a function of $\rho$, one finds
\bqa
\label{inrho1}
\mu_{0+1}(\rho)=g\rho+{g^2\rho\over8\pi}\bigg[\bar{L}+1\bigg]\;,
\eqa 
where $\bar{L}=\log\left({g\rho/2M^2}\right)$
and $g=g(M)$.
Using Eqs.~(\ref{rela}),~(\ref{f1}) and~(\ref{inrho1}), 
the energy density in the one-loop approximation becomes
\bqa
\label{e1}
{\cal E}_{0+1}(\rho)={1\over2}g\rho^2
+{g^2\rho^2\over16\pi}\bigg[\bar{L}+{1\over2}\bigg]\;.
\eqa
This agrees with the result obtained 
by Lozano~\cite{loz}, and by Haugset and 
Ravndal~\cite{finn}. 
\subsection{Two-loop energy density}
Using Eqs.~(\ref{rhodef}) and~(\ref{f012}), we obtain the density in the
two-loop approximation:
\bqa
\rho_{0+1+2}(\mu)&=&{\mu\over g}
-{\mu\over8\pi}\bigg[L+1\bigg]
+{g\mu\over32\pi^2}\left[1+C\right]
\;.
\label{rho2}
\eqa
Inverting Eq.~(\ref{rho2}), we obtain the two-loop expression for the
chemical potential:
\bqa\nonumber
\mu_{0+1+2}(\rho)&=&g\rho+{g^2\rho\over8\pi}\bigg[\bar{L}+1\bigg]
\\
&&
+{g^3\rho\over64\pi^2}\left[\bar{L}^2+3\bar{L}-2C\right]
\;.
\eqa 
The final result for the two-loop energy density then becomes
\bqa\nonumber
{\cal E}_{0+1+2}(\rho)&=&{1\over2}g\rho^2
+{g^2\over16\pi}\rho^2\left[\bar{L}+{1\over2}\right]
\\
&&
+{g^3\over128\pi^2}\rho^2\left[\bar{L}^2+
2\bar{L}-1-2C\right]\;.
\label{final}
\eqa
Using the renormalization group equation~(\ref{rg2}) 
for the
running coupling constant, we see that the two-loop results for the
density, chemical potential,
and energy density are independent
of the renormalization scale $M$ up to corrections of order $g^4$. 
\section{Renormalization group}
In Secs.~III and IV, we showed that the results 
for the pressure, density, chemical potential, and
energy density are independent of the arbitrary renormalization
scale $M$ that is introduced with dimensional regularization.

The two-loop results for the pressure and density include a logarithm
$L$.
The renormalization group can be used to absorb
this logarithm.
The solution to the renormalization group equation~(\ref{rg2}) is
\bqa
\label{sol}
g(M_0)={g(M)\over{1-{g(M)\over8\pi}\log{\left(M_0^2\over M^2\right)}}}\;,
\eqa
By substituting the running coupling constant into the expressions for the
pressure and density in the mean-field approximation
and choosing the renormalization scale $M_0=\sqrt{\mu/2}$, we 
absorb this logarithm.
In a similar manner, we can sum logs of the form
$g^{n+1}L^n$ ($n=1,2,3...$) by using the running coupling in the
remaining terms in expressions for the pressure and density.
These terms are generated by expanding out Eq.~(\ref{sol}) in powers of ${L}$. 
The result is
\bqa
\rho(\mu)
&=&{\mu\over g_0}-{\mu\over8\pi}+{g_0\mu\over32\pi^2}\left[1+C\right] \;,
\\
{\cal P}(\mu)&=&
{\mu^2\over2g_0}-{\mu^2\over32\pi}+{g_0\mu\over64\pi^2}\left[1+C\right] \;,
\eqa
where  $g_0\equiv g(M_0)$.
In the same way, we can sum leading (of the form 
$g^{n+1}\bar{L}^n$, where $n=1,2,3...$) and subleading logs
(of the form 
$g^{n+m}\bar{L}^{n}$, where $n=1,2,3...$, and $m=2,3...$,)
from all orders of perturbation theory 
in the chemical potential and energy density
by using a running coupling constant 
$g(M_0)$ and choosing the renormalization scale $M_0=\sqrt{\rho/2}$.
These terms are generated by expanding out Eq.~(\ref{sol}).
The result is
\bqa\nonumber
\mu(\rho)&=&g_0\rho+{g^2_0\rho\over8\pi}\left[\log(g_0)+1\right]
\\
&&
+{g^3_0\rho\over64\pi^2}
\left[\log^2(g_0)+3\log(g_0)-2C\right]\;,\\ \nonumber
{\cal E}(\rho)&=&
{1\over2}g_0\rho+{g^2_0\rho\over16\pi}\left[\log(g_0)+{1\over2}\right]
\\
&&
+{g^3_0\rho\over128\pi^2}
\left[\log^2(g_0)+2\log(g_0)-1-2C\right]\;,
\eqa
where $g_0\equiv g(M_0)$.

We have approximated the two-body potential by local interactions.
In real systems, however,  the interactions have a finite range $a$ and
so $1/a$ provides a natural ultraviolet cutoff $M$. 
In the dilute gas limit, we have
$|\log(\mu a^2/2)|\gg1$ and $|\log(\rho a^2/2)|\gg1$
~\cite{schiff,popov1,popov2,fis,rg2,fried},
and the running coupling constant
is approximately equal to either
$-8\pi[\log(\mu a^2/2)]^{-1}$
or $-8\pi[\log(\rho a^2/2)]^{-1}$.
This is essentially the two-body $T$-matrix and so this is the relevant
expansion parameter.
By substituting these
expressions for the coupling constant into to the expressions
for the pressure, density, chemical potential, and energy density, we 
obtain the renormalization group improved results
\bqa
\label{pfinal}
{\cal P}(\mu)&=&-{\mu^2L\over16\pi}
\left[1+{1\over2}L^{-1}+2(1+C)L^{-2}\right]\;, \\
\rho(\mu)&=&-{\mu L\over8\pi}
\left[1+L^{-1}+2(1+C)L^{-2}\right]\;,
\\
\nonumber
\mu(\rho)&=&-{8\pi\rho\over\bar{L}}
\left[
1+
\left(\log[-\bar{L}/8\pi]-1\right)\bar{L}^{-1}
\right. \\ \nonumber
&&
+\log^2\left(\log[-\bar{L}/8\pi]\right)L^{-2}
-{3}\left(\log[-\bar{L}/8\pi]\right)L^{-2}
\\
&&\left.
-2C\bar{L}^{-2}
\right]
\label{myfinal}
 \;, \\
\nonumber
{\cal E}(\rho)&=&-{4\pi\rho^2\over\bar{L}}
\left[
1+{1\over2}\left(2\log[-\bar{L}/8\pi]-1\right)\bar{L}^{-1}
\right.\nonumber
\\ \nonumber
&&
+\log^2\left(\log[-\bar{L}/8\pi]\right)L^{-2}
-{2}\left(\log[-\bar{L}/8\pi]\right)L^{-2}
\\
&&
\left.
-(1+2C)\bar{L}^{-2}
\right]
\label{efinal}
\;,
\eqa
where where $L=\left[\log({\mu a^2}/2)\right].$
and
$\bar{L}=\left[\log(\rho a^2/2)\right]$.
Eqs.~(\ref{pfinal})-(\ref{efinal}) are the main result of the 
present paper. The 
leading order results 
were first obtained by Schick~\cite{schiff}, while 
the leading corrections have been calculated
in Refs.~\cite{popov1,popov2,fis,rg2,fried,hines}.
The last terms are new result.
\section{Summary}
In the present paper, we have studied a two-dimensional interacting 
homogeneous Bose gas
at zero temperature. We have calculated 
the ground state pressure and energy density to second order in the
quantum loop expansion. The results are independent of the
arbitrary renormalization scale $M$.

We have applied the renormalization group to sum up leading and subleading
logarithms from all orders in perturbation theory. 
The renormalization group improved pressure and density are 
expansions in $\left[\log(\mu a^2/2)\right]^{-1}$,
while the chemical potential and energy density are expansions in 
$\left[\log(\rho a^2/2)\right]^{-1}$.
These are essentially expansions in the two-body $T$-matrix,
and we have obtained the ground state pressure, density, chemical
potential, and energy density 
to next-to-next-to-leading order.

\section*{Acknowledgments}
This work was supported by the Stichting voor
Fundamenteel Onderzoek der Materie
(FOM), which is supported by the Nederlandse Organisatie voor Wetenschapplijk
Onderzoek (NWO). 
\appendix
\renewcommand{\theequation}{\thesection.\arabic{equation}}
\setcounter{equation}{0}
\section{Formulas}
The loop integrals that appear in our calculations involve integrations
over the energy $\omega$ and the spatial momentum ${\bf p}$.
The energy integrals are evaluated using contour integration. 

The specific one-loop integral needed is
\bqa
\label{e11}
\int{d\omega\over2\pi}{1\over[\omega^2-\epsilon^2(p)+i\epsilon]}&=&
-{i\over2\epsilon(p)}\;.
\eqa
The specific two-loop integrals needed are
\bqa
\nonumber
\int{d\omega_1\over2\pi}\int{d\omega_2\over2\pi}
{1\over\left[\omega_1^2-\epsilon^2(p)+i\epsilon\right]}
\times
&&\\ \nonumber
{1\over[\omega_2^2-\epsilon^2(q)+i\epsilon][(\omega_1+\omega_2)^2-\epsilon^2(r)+i\epsilon]}
&=&\\
{1\over4\epsilon(p)\epsilon(q)\epsilon(r)(\epsilon(p)+\epsilon(q)+
\epsilon(r))}\;,\\ \nonumber
\int{d\omega_1\over2\pi}\int{d\omega_2\over2\pi}
{\omega_1\omega_2\over[\omega_1^2-\epsilon^2(p)+i\epsilon]}
\times
&& \\ \nonumber
{1\over[\omega_2^2-\epsilon^2(q)+i\epsilon][(\omega_1+\omega_2)^2-\epsilon^2(r)+i\epsilon]}
&=&
\\
{1\over4\epsilon(r)[\epsilon(p)+\epsilon(q)+
\epsilon(r)]}\;.
&&
\eqa
Here $r=|{\bf p}+{\bf q}|$.

Some of the one-loop momentum
integrals are infrared divergent or ultraviolet divergent or
both.
They can be written in terms of the integral $I_{m,n}$, which is defined by
\bqa
\label{idef}
I_{m,n}=\left({e^{\gamma}M^2\over4\pi}\right)^{\epsilon}
\int{d^dp\over(2\pi)^d}{p^{2m}\over p^n(p^2+\Lambda^2)^{n/2}}\;.
\eqa
Here, $M$ is a renormalization scale that ensures that $I_{m,n}$ has the
canonical dimension also for $d\neq2$.
$\gamma\approx0.5772$ is the Euler-Mascheroni constant. 
With dimensional regularization, $I_{m,n}$ is given by the formula
\bqa\nonumber
I_{m,n}&=&{\Omega_d\over(2\pi)^d}
\left({e^{\gamma}M^2\over4\pi}\right)^{\epsilon}
\Lambda^{d+2m-2n}
\\
&&
{\Gamma({{d-n\over2}+m})\Gamma(n-m-{d\over2})\over2\Gamma({n\over2})}\;,
\eqa
where $\Omega_d=2\pi^{d/2}/\Gamma[d/2]$ is the 
area of the $d$-dimensional sphere.

The integrals $I_{m,n}$ satisfy the relations
\bqa
\label{alge}
{\!\!\!d\over d\Lambda^2}I_{m,n}&=&-{n\over2}I_{m+1,n+2}\;,\\
\left(d+2m-n\right)I_{m,n}&=&nI_{m+2,n+2}\;,\\
\Lambda^2I_{m,n}&=&I_{m-1,n-2}-I_{m+1,n}\;.
\eqa
The first relation follows directly from the definition of $I_{m,n}$.
The second relation follows
from integration by parts, while the last
is simply an algebraic relation.

In two dimensions, these integrals have logarithmic and power ultraviolet
divergences. The power divergences are set to zero in dimensional 
regularization, while the logarithmic divergences appear as poles in 
$\epsilon$. The specific integrals are
\bqa\nonumber
I_{0,-1}&=&-{\Lambda^4\over32\pi}
\left\{{1\over\epsilon}-L
-{1\over2}
\right.\\
&&
\left.
+{1\over2}\left[
L^2+L-{5\over2}+{\pi^2\over2}
\right]
\epsilon+
{\cal O}\left(\epsilon^2\right)
\right\}
\label{i1}
\;,\\ \nonumber
I_{-1,-1}&=&{\Lambda^2\over8\pi}\left\{{1\over\epsilon}
-L+1
\right.\\
&&
\left.
+{1\over2}\left[
\left(L-1\right)^2
+1+{\pi^2\over2}\right]\epsilon
+{\cal O}\left(\epsilon^2\right)
\right\}\;,\\
\label{i3}\nonumber
I_{1,1}&=&-{\Lambda^2\over8\pi}\left\{{1\over\epsilon}
-L-1
\right.\\ 
&&
\left.
+
{1\over2}\left[
(L+1)^2-4+{\pi^2\over2}
\right]\epsilon
+{\cal O}\left(\epsilon^2\right)
\right\}
\label{i2}
\;, \\
I_{0,1}&=&{1\over4\pi}\left\{{1\over\epsilon}
-L
+{1\over2}\left[
L^2
+{\pi^2\over2}
\right]\epsilon
+{\cal O}\left(\epsilon^2\right)
\right\}
\;,
\label{i4}
\eqa
where $L=\log(\Lambda^2/4M^2)$.

The two-loop integrals needed can be expressed in terms of $J_{l,m,n}$
\bqa\nonumber
\label{jlmn}
J_{l,m,n}&=&\left({e^{\gamma}M^2\over4\pi}\right)^{2\epsilon}
\int{d^dp\over(2\pi)^d}\int{d^dq\over(2\pi)^d}
\times
\\
&&
\hspace{-1.4cm}
{\Big[p/\sqrt{p^2+\Lambda^2}\Big]^{l}
\Big[q/\sqrt{q^2+\Lambda^2}\Big]^{m}
\Big[r/\sqrt{r^2+\Lambda^2}\Big]^{n}
\over p\sqrt{p^2+\Lambda^2}+q\sqrt{q^2+\Lambda^2}+r\sqrt{r^2+\Lambda^2}}
\;.
\eqa 
In two dimensions, these integrals have quadratic and double logarithmic
divergences that cancel in the particular combination
in Eq.~(\ref{f2j}), 
leaving us with a logarithmically divergent integral.
We write the integral as
\bqa
J=J_{\rm div}+J_{\rm num}\;,
\eqa
where 
the ultraviolet divergence of 
the integral $J$ has been isolated:
\bqa\nonumber
J_{\rm div}&=&2\left({e^{\gamma}M^2\over4\pi}\right)^{2\epsilon}
\int{d^dp\over(2\pi)^d}
\left[2-{p\over\sqrt{p^2+\Lambda^2}}
\right.\\&&
\left.
-{\sqrt{p^2+\Lambda^2}\over p}
\right]
\int{d^dq\over(2\pi)^d}{1\over q\sqrt{q^2+\Lambda^2}}
\label{div}
\;.
\eqa
The first term inside the square brackets in Eq.~(\ref{div})
has only a power divergence, and so it is set to zero in 
dimensional regularization. In terms of $I_{m,n}$, the remaining terms
can be written as
\bqa
J_{\rm div}=-2I_{0,1}\left[
I_{1,1}+I_{-1,-1}
\right]\;.
\eqa
The remaining finite part of the integral $J$ can be evaluated directly
in two dimensions and reads
\bqa\nonumber
J_{\rm num}&=&
\int{d^2p\over(2\pi)^2}\int{d^2q\over(2\pi)^2}\Bigg\{
\left[{6p\over \sqrt{p^2+\Lambda^2}}
-{2\over p}\sqrt{p^2+\Lambda^2}
\right.\\ \nonumber
&&
-{3pqr\over \sqrt{p^2+\Lambda^2}\sqrt{q^2+\Lambda^2}\sqrt{r^2+\Lambda^2}}
\\ \nonumber
&&
\left.
-{r\sqrt{p^2+\Lambda^2}\sqrt{q^2+\Lambda^2}\over
pq\sqrt{r^2+\Lambda^2}}\right]
\times
\\ \nonumber
&&
{1\over p\sqrt{p^2+\Lambda^2}+q\sqrt{q^2+\Lambda^2}
+r\sqrt{r^2+\Lambda^2}}
\\\nonumber
&&
-{2\over q\sqrt{q^2+\Lambda^2}}\left[
2
-{p\over\sqrt{p^2+\Lambda^2}}
\right.
\\
&&
\left.
-{\sqrt{p^2+\Lambda^2}\over p}
\right]
\Bigg\}
\;.
\label{jint}
\eqa
Since the only scale in the integrand in Eq.~(\ref{jint}) is $\Lambda$,
it follows from dimensional analysis that
$J_{\rm  num}$ is proportional to $\Lambda^2$. The numerical value is
\bqa
J_{\rm num}=-1.76\times10^{-5}\Lambda^2\;.
\eqa

\end{multicols}

\begin{thebibliography}{99}
\bibitem{bec1} M.H. Anderson, J.R. Ensher, M.R. Matthews, C.E. Wieman
and E.A. Cornell, Science {\bf 269}, 198 (1995).
\bibitem{bec2}
K.B. Davis, M.O. Mewes, M.R. Andrews, N.J. van Druten,
D.S. Durfee, D.M. Kurn and W. Ketterle, Phys. Rev. Lett. {\bf 75}, 
3969 (1995).
\bibitem{bec3}
C.C. Bradley, C.A. Sackett, J.J. Tollett and R.G. Hulet, 
Phys. Rev. Lett. {\bf 75}, 1687 (1995).

\bibitem{string} F. Dalfovo, S. Giorgini, L.P. Pitaevskii, and S. Stringari, Rev. Mod. Phys. {\bf 71}, 463 (1999).
\bibitem{grif}H. Shi and A. Griffin, Phys. Rep. {\bf 304}, 1 (1998).

\bibitem{leeyang}T.D. Lee and C.N. Yang, Phys. Rev. {\bf 105}, 1119 (1957).
\bibitem{to} T.T. Wu, Phys. Rev. {\bf 115}, 1390 (1959); N.M. Hugenholz
and D. Pines, Phys. Rev. {\bf 116}, 489 (1959); K. Sawada, 
Phys. Rev. {\bf 116}, 1344 (1959).
\bibitem{eric}E. Braaten and A. Nieto, Eur. J. Phys, (1997).
\bibitem{e2}E. Braaten, H.-W. Hammer and S. Hermans, Phys. Rev. {\bf A63}, 063609 (2001). 

\bibitem{schiff}M. Schick, Phys. Rev. {\bf A3}, 1067 (1971).

\bibitem{popov1} V.N. Popov, Theor. Math. Phys. {\bf 11}, 565 (1972).
\bibitem{popov2} V.N. Popov {\it Functional Integrals in Quantum Field Theory and Statistical Physics}, (Reidel, Dordrecht, 1983). 
\bibitem{fis}D.S. Fisher and P.C. Hohenberg, Phys. Rev. {\bf B37}, 4936 (1988).\bibitem{rg2}E.B. Kolomeisky and J.P. Straley, Phys. Rev. {\bf B46}, 11749 (1992).
\bibitem{fried} C.-C. Chang and R. Friedberg, Phys. Rev. {\bf B51}, 1117 (1995).
\bibitem{hines} D.F. Hines, N.E. Frankel, and D.J. Mitchell, Phys. Lett. {\bf A68}, 12 (1978).
\bibitem{lieb1}E.H. Lieb and J. Yngvason, e-print math-ph/0002014.
\bibitem{lieb2}E.H. Lieb, R. Seiringer, and J. Yngvason, e-print 
cond-mat/0005026.


\bibitem{berg}O. Bergman, Phys. Rev. {\bf D46}, 5474 (1992). 
\bibitem{loz}G. Lozano, Phys. Lett. {\bf B283}, 70 (1992).
\bibitem{finn}T. Haugset and F. Ravndal, Phys. Rev. {\bf D49}, 4299 (1994).


\end{thebibliography}
\end{document}